\documentclass[12pt]{article}
\def\lsim{\mathrel{\lower2.5pt\vbox{\lineskip=0pt\baselineskip=0pt
          \hbox{$<$}\hbox{$\sim$}}}}
\def\gsim{\mathrel{\lower2.5pt\vbox{\lineskip=0pt\baselineskip=0pt
          \hbox{$>$}\hbox{$\sim$}}}}
\renewcommand{\theequation}{\arabic{section}.\arabic{equation}}
\begin{document}
%----------------------------------------------------------------------------- 
\markright{Zeta Functions and the Casimir Energy}
%----------------------------------------------------------------------------
% Preprint: LA-UR-88-1542 
% Preprint: MPI-PAE/PTh-18/88
%----------------------------------------------------------------------------
\title{ {\bf Zeta Functions and the Casimir Energy} }
%----------------------------------------------------------------------------
\author{Steven K. Blau and Matt Visser\\[2mm]
\small\it Theoretical Division \\
\small\it T8, Mail Stop B-285 \\
\small\it Los Alamos National Laboratory\\
\small\it Los Alamos, New Mexico 87545, USA\\[8mm]
Andreas Wipf\\[2mm]
\small\it Max Planck Institut f\"{u}r Physik und Astrophysik \\
\small\it ---Werner Heisenberg Institut f\"{u}r Physik---\\
\small\it Postfach 40 12 12
\small\it D-8000 M\"{u}nchen 40, FRG}
%-----------------------------------------------------------------------------
\date{{\small 24 May 1988; arXiv-ed 15 June 2009; \LaTeX-ed \today}}
%------------------------------------------------------------------------------
\maketitle
%------------------------------------------------------------------------------
\begin{abstract}
We use zeta function techniques to give a finite definition for the
Casimir energy of an arbitrary ultrastatic spacetime with or without
boundaries.  We find that the Casimir energy is intimately related to,
but not identical to, the one--loop effective energy.  We show that in
general the Casimir energy depends on a normalization scale.  This
phenomenon has relevance to applications of the Casimir energy in bag
models of QCD.

Within the framework of Kaluza--Klein theories we discuss the
one--loop corrections to the induced cosmological and Newton constants
in terms of a Casimir like effect. We can calculate the dependence of
these constants on the radius of the compact dimensions, without
having to resort to detailed calculations.
 
\vspace*{5mm}
\noindent
%PACS: 
%\\
Keywords: Zeta functions, Casimir energy, effective energy, effective action.

\bigskip
%-----------------------------------------------------------------------
\centerline{\underline{Published as:} Nuclear Physics {\bf B310} (1988) 163--180.}
%----------------------------------------------------------------------
\medskip
%----------------------------------------------------------------------
\centerline{\underline{doi:} 10.1016/0550-3213(88)90059-4   (abstract only).   }
%----------------------------------------------------------------------

\end{abstract}
%-----------------------------------------------------------------------
\bigskip
%----------------------------------------------------------------------
\hrule
%-----------------------------------------------------------------------
\bigskip
%-----------------------------------------------------------------------
%-----------------------------------------------------------------------
\centerline{{E-mail (2009):} {\sf sblau@aip.org, matt.visser@msor.vuw.ac.nz,  wipf@tpi.uni-jena.de}}
%-----------------------------------------------------------------------
\bigskip
%-----------------------------------------------------------------------
%-----------------------------------------------------------------------
\centerline{{URL (2009):} {\sf http://homepages.msor.vuw.ac.nz/$\sim$visser/  \quad http://www2.uni-jena.de/$\sim$p5anwi/ }}
%-----------------------------------------------------------------------
\bigskip
\hrule
\clearpage
%---------------------------------------------------------------------
% User definitions
%--------------------------------------------------------------------
\def\Box{\nabla^2}
%---------------------------
\def\d{{\mathrm d}}
%----------------------------
\def\ie{{\em i.e.\/}}
\def\eg{{\em e.g.\/}}
\def\etc{{\em etc.\/}}
\def\etal{{\em et al.\/}}
%----------------------------
\def\S{{\mathcal S}}
\def\I{{\mathcal I}}
\def\L{{\mathcal L}}
\def\R{{\mathcal R}}
\def\M{{\mathcal M}}
\def\H{{\mathcal H}}
%-----------------------------
\def\tr{{\mathrm{tr}}}
\def\implies{\Rightarrow}
\def\half{{1\over2}}
\def\quarter{{1\over4}}
%------------------------------
\def\normal{{\mathrm{normal}}}
\def\induced{{\mathrm{induced}}}
\def\stringy{{\mathrm{stringy}}}
\def\graviton{{\mathrm{graviton}}}
\def\critical{{\mathrm{critical}}}
\def\eff{{\mathrm{eff}}}
\def\surface{{\mathrm{surface}}}
\def\bulk{{\mathrm{bulk}}}
\def\boundary{{\mathrm{boundary}}}
\def\matter{{\mathrm{matter}}}
%---------------------------------
\def\Newton{{\mathrm{Newton}}}
\def\Planck{{\mathrm{Planck}}}
%------------------------------------
%-----------------------------------------------------------------------
% definitions peculiar to this paper
%-------------------------------------------------------------------------------
\def\z{\zeta}
\def\Bigrho{\rho}
\def\Bigsigma{\sigma}
\def\curl{\partial}
%---------------------------------
\def\e{ {E_{\rm Casimir}} }
\def\Ec{ E_{\rm Casimir} }
\def\Er{ E_{\rm reg} }
\def\Es{ {E_{\rm subtracted}} }
%---------------------------------
\def\t{\textstyle}
\def\s{\scriptstyle}
%--------------------------------
\def\dslash#1{{#1\!\!\!\slash\,}}
\def\gab{g_{\alpha\beta}}
\def\bd{{\curl\Omega}}                      %boundary
%--------------------------------
\def\Seff{{S_{\eff}}}
\def\Leff{{L_{\eff}}}
\def\Eeff{{E_{\eff}}}
%--------------------------------
\def\A{{\s A}}  %\s=\scriptstyle!
\def\B{{\s B}}
\def\C{{\s C}}
\def\D{{\s D}}
\def\E{{\s E}}
\def\F{{\s F}}
%-------------------------------
\def\L{{\cal L}}
\def\I{{\cal I}}
\def\M{{\cal M}}
%-------------------------------
\def\Ricci{\hbox{\em Ricci}}
\def\Weyl{\hbox{\em Weyl}}
%---------------------------------------------------------------
\def\dirac{{D\!\!\!\!\slash\,}}
\def\Dirac{{\nabla\!\!\!\!\slash\,}}
%---------------------------------------------------------------
\def\eps{\epsilon}
%---------------------------------------------------------------
\def\Break{\hfil\break}
\def\ss{\scriptscriptstyle}
\def\s{\scriptstyle}
\def\abs#1{\vert #1 \vert}
\def\Sum{\sum}
\def\script{\cal}  %normally cmsy10 font!
%--------------------------------------------------------------- 
\def\order{o}
%---------------------------------------------------------------
\def\infinity{\infty}
%---------------------------------------------------------------
\tableofcontents
\bigskip
\hrule
%--------------------------------------------------------------- 
\section{\bf Introduction.}
%--------------------------------------------------------------- 
\setcounter{equation}{0}
%---------------------------------------------------------------------

The study of vacuum fluctuations, as embodied in the Casimir
effect~\cite{Casimir}, has been a subject of extensive
research~\cite{Review:Casimir}.  The Casimir energy may be thought of
as the energy due to the distortion of the vacuum.  This distortion
may be caused either by some background field (\eg\ gravity), or by
the presence of boundaries in the space--time manifold (\eg\
conductors).  Early investigations of the effects of a gravitational
background were performed by Utiyama and De Witt~\cite{U-DeWitt}, and
work has continued on this important
subject~\cite{Dowker-Critchley,Ford,HaKi,Deutsch-Candelas,Kay,Dowker}.
Early work on the effect of boundaries was performed by
Casimir~\cite{Casimir}, and was later extended by Fierz, Boyer,
deRaad, and Milton~\cite{Fierz,Boyer,Milton-On-Balls,deRaad-Milton}.
More recently boundary effects have been central to the calculation of
the Casimir energy in bag models of
QCD~\cite{Bender-Hays,Milton,Baacke}.
 
We feel that interesting things remain to be said.  In this paper heat
kernel and zeta function techniques will be utilized to investigate
these topics~\cite{Dowker-Critchley,Hawking}.  The unified treatment
presented here is applicable to a very wide class of models and
physical situations.
 
We start by developing a definition of the Casimir energy which is
{\em finite} and applies to arbitrary static manifolds with or without
boundaries
\begin{equation}
\Ec = \half\hbar c\mu\cdot PP[\z_3(-{\t{1\over2}}+\eps)].
\end{equation}
Here $\mu$ is a normalization scale of dimension $(length)^{-1}$, and
the PP symbol indicates that we are to extract the ``principal part''.
This definition yields a finite quantity in both flat and curved
space--times, with or without boundaries, for both massive and
massless particles. The normalization scale $\mu$ appearing in the
above is required to keep the zeta function dimensionless for all
values of $s$. The introduction of this scale leads generically to
non-trivial scaling behaviour for the Casimir energy. It is pointed
out how this definition relates in special cases to well--known
results.
 
Our definition of the Casimir energy allows us to investigate its
dependence on the ``radius'' of the manifold. We find that for
massless fields
\begin{equation}
\Ec(R) = {\hbar c\over R} \cdot\{\eps_0 - \eps_1\cdot\ln(\mu R) \},
\label{E:radius}
\end{equation}
where the $\mu$-independent coefficients $\eps_0$ and $\eps_1$ are
dimensionless numbers depending on the geometry of the manifold. This
result has some very interesting consequences when applied to the bag
models of hadrons in QCD.
 
Further, we may relate the Casimir energy to the one--loop effective
action (\ie\ the determinant of a suitable {\em four} dimensional
differential operator). This is done by relating the zeta function of
$D_4 = -\curl_0{}^2 + D_3$ to the zeta function of $D_3$
\begin{equation}
\z_4(s) = {\mu c T\over\sqrt{4\pi}}\cdot{\Gamma(s-{1\over2})\over\Gamma(s)}
                \cdot\z_3(s-{\t{1\over2}}).
\end{equation}
Thus we obtain a non-trivial relationship between the
Casimir energy and the one-loop effective energy
\begin{equation}
E_\eff = 
\e + \half\hbar c\mu \; \left[\psi(1)-\psi(-\t{\half})\right]\;
{C_2\over(4\pi)^2}.
\end{equation}
To help understand the significance of this relationship we include a
discussion of the various {\em different} concepts commonly lumped
together as ``vacuum energy''.
 
We next apply our analysis to the one--loop corrections to the
effective cosmological constant and Newton constant in Kaluza--Klein
theories. These one-loop corrections may be interpreted as a
Casimir-like effect. We derive the following {\em finite} expressions
for the one-loop four-dimensional effective cosmological and Newton
constants.
\begin{eqnarray}
\Lambda_{\rm eff} &=& 
\Lambda\cdot {\mathrm{vol}} (\Omega) + 
G^{-1}\cdot\int_\Omega \sqrt{g} \; R_d -
{\mu^4\over2(4\pi)^2} \;\left\{ {\t\half}\z'_d(-2) - {\t{3\over4}}\z_d(-2) \right\}, 
\nonumber\\
G_{\rm eff}^{-1} &=& 
G^{-1}\cdot {\mathrm{vol}} (\Omega)  -
k \;{\mu^2\over2(4\pi)^2}\; \left\{ \z'_d(-1)  - \z_d(-1) \right\}. 
\end{eqnarray}
In particular, this allows us to study the dependence of these
constants on the ``radius'' of the compact dimensions, without having
to resort to explicit calculations.
 
%---------------------------------------------------------------
%\clearpage
%---------------------------------------------------------------
\section{\bf Zeta functions on manifolds with boundary.}
%--------------------------------------------------------------- 
\setcounter{equation}{0}
%---------------------------------------------------------------------

As regularization technique we shall use the zeta-function method due
to Dowker and Critchley~\cite{Dowker-Critchley} and
Hawking~\cite{Hawking}. Its relation to other methods (\eg,
dimensional regularization) has been discussed in the
literature~\cite{Dowker-Critchley}.  In order to make subsequent
arguments understandable, we must first briefly review the
mathematical machinery of zeta functions. Consider the zeta function
associated with a second-order self-adjoint elliptic operator $D$
defined on a compact manifold $\Omega$ with boundary $\bd$
\begin{equation}
\z(s)  = \tr'\{(\mu^{-2} D)^{-s}\}  = {\sum}' (\mu^{-2}\lambda_n)^{-s},
\end{equation}
where the $\lambda_n$ are eigenvalues of $D$; while the prime on
$\tr'$ and $\sum'$ indicates that we should {\em not} include the zero
eigenvalues of $D$ in the sum. We have introduced a ``scale'' $\mu$,
with the dimensions of $(length)^{-1}$, in order to keep the
zeta-function dimensionless for all $s$.
 
The zeta function is related to the diffusion operator (heat kernel)
via a Mellin transform:
\begin{eqnarray}
\z(s) &=& {\sum}'{1\over\Gamma(s)} \int_0^\infinity dt \; t^{s-1} \;
                \exp(-\lambda_n \mu^{-2} t) 
\nonumber\\
      &=& {1\over\Gamma(s)}\int_0^\infinity dt \;  t^{s-1} \;
                \tr'(e^{-t D \mu^{-2}}). 
\label{E:Mellin}
\end{eqnarray}
Here $t$ is a dimensionless parameter, not to be confused with
physical time ($x^0/c$). From now on, in the interests of notational
simplicity, we ignore zero modes. The trace of the diffusion operator
is given by the integral of the diagonal part of the heat kernel over
the manifold:
\begin{equation}
\tr(e^{-t D \mu^{-2}}) = \int_\Omega K(t,x,x)\; \sqrt{g}\; d^dx.
\end{equation}
The heat kernel $K$ possesses an asymptotic expansion for small $t$:
\begin{equation} 
\label{E:an}
K(t,x,x)=\left({\mu^2\over4\pi t}\right)^{d/2} \cdot
      \left\{ \sum_0^N     a_n(x) \; (\mu^{-2}t)^n+ \order(t^N) \right\}.
\end{equation}
The sum is over integer values of $n$.  The $a_n$ are functions of the
gravitational field, they may be expressed as polynomials in the
Riemann tensor, its contractions, and covariant derivatives. (See
Appendix A.) The diagonal part of the heat kernel contains
exponentially suppressed terms ($e^{-k(x)/t}$) that do not contribute
to the asymptotic expansion (\ref{E:an}). These exponentially
suppressed terms do however contribute an explicit boundary term to
the trace of the heat kernel
\begin{equation}
\label{E:bn}
\tr(e^{-t D \mu^{-2}})=\left({\mu^2\over4\pi t}\right)^{d/2} \cdot
      \left\{ \sum_0^N  \left( \int_\Omega a_n(x) \; (\mu^{-2}t)^n
                + \int_\bd    b_n(y) \; (\mu^{-2}t)^n  \right)
                + \order(t^N) \right\}.
\end{equation}
The sum runs over half--integers, (but the $a_n$ vanish for
half-odd-integers).  The $b_n$ are functions of the second fundamental
form of the boundary (extrinsic curvature), the induced geometry on
the boundary (intrinsic curvature), and the nature of boundary
conditions imposed. These objects are tabulated in many places: \eg,
Birrell and Davies~\cite{BD} and Appendix A of this paper. For future
reference we define the dimensionless quantities: $A_n = \mu^{d-2n} \;
 \int_\Omega a_n(x)\; \sqrt{g} \; d^dx$, $B_n = \mu^{d-2n} \;
\int_\bd \; b_n(y) \; \sqrt{\tilde g} \,d^{d-1}y $, and $C_n = A_n + B_n$.
 
In view of the asymptotic expansion (\ref{E:bn}), it is clear that the
zeta function $\z(s)$ is a meromorphic function of the complex
variable $s$ possessing only simple poles whose residues are
determined by $C_n$. Observe that (\ref{E:bn}) implies that $\z(s)$
has a pole structure given by
\begin{equation}
\z(s) = {1\over\Gamma(s)\;(4\pi)^{d/2}} \cdot \left\{
         \sum_0^\infinity {C_n\over(s-[{d\over2}-n])}
         + f(s) \right\}.
\end{equation}
 
The function $f(s)$ is an entire analytic function of $s$, but, in
general, we have little additional information concerning its
behaviour.  However, we do know that $\z(s)$ is analytic at $s=0$. It is
thus possible to define the determinant of $D$ to be~\cite{Hawking}
\begin{equation}
{\det}'(\mu^{-2} D) = \exp\left( - \left.{d\over ds} \z(s) \right\vert_{s=0}
\right).
\end{equation}
Observe that many of the technical details associated with
renormalization have been hidden by these zeta function techniques.
We shall now utilize this mathematical machinery to define the Casimir
energy, and relate $\Ec$ to the one--loop Effective action
$\Seff=\half\ln\det D$.

%\clearpage

%---------------------------------------------------------------
\section{\bf The Casimir energy.}
%--------------------------------------------------------------- 
\setcounter{equation}{0}
%---------------------------------------------------------------------

In order to have a well--defined notion of energy, it is useful to
work in a static spacetime~\cite{DeWitt}, specifically let us take
$g_4 = -(dx^0)^2 + g_3$, in which case we decompose the differential
operator $D_4$ as $D_4 = -(\curl_0)^2 + D_3$.  The eigen-frequencies
associated with $D_3$ are $\omega_n = \sqrt{\lambda_n(D_3)}\cdot c$.
We wish to consider the zero--point energy:
\begin{equation}
\Ec = {\t\half} \sum_n \hbar \omega_n .
\end{equation}
This sum is, of course, divergent. We regularize it by defining
\begin{equation}
\Er(\eps) = \t\half \hbar c\mu \cdot \sum_n
             (\lambda_n\mu^{-2})^{(\half - \eps)}
          = \t\half \hbar c\mu \cdot \z_3\left(-\t\half+\eps\right).
\label{E:regulated}
\end{equation}
Where $\z_3$ is the zeta function associated with the
three-dimensional operator $D_3$. A quick glance at the previous
section shows that $\Er(\eps)$ is a meromorphic function with a pole
at $\eps=0$, with residue $-\half\hbar c\mu \; C_2(g_3)/(4\pi)^2 =
-\half\hbar c \{ \int_{\Omega} a_2 + \int_{\bd} b_2 \}/(4\pi)^2$,
where the integral is over three-dimensional space and its
two-dimensional boundary. Because of the pole at $\eps=0$, we cannot,
in general, remove the regulator; the geometric coefficient $C_2$ is
an obstacle to giving a finite definition for the Casimir energy.
Note, however, that in many interesting cases (\eg, flat space with
flat boundaries and massless particles) $C_2=0$, so that
$\lim_{\eps\to0} \Er(\eps)$ is finite, and independent of the
normalization scale $\mu$.

How is one to understand the unphysical pole and $\mu$ dependence of
the (zeta-function regulated) Casimir energy? First we note that the
Casimir energy in isolation is unphysical. When physicists speak of
the Casimir energy they usually are identifying terms in the
renormalized total energy which they interpret as arising from boundary
or gravitational effects. There is {\em ipso facto} no pole in the
total energy; the pole in equation (\ref{E:regulated}) is absorbed
into the bare action which must contain a term proportional to
$C_2$. Having seen this we must admit that the way in which the pole
is removed is not unique. The possibility of different renormalization
schemes means that the Casimir energy has an ambiguity proportional to
$C_2$. Our choice of renormalization scheme is to adopt the minimal
subtraction scheme which is equivalent to simply removing the pole
from equation (\ref{E:regulated}). We define
\begin{eqnarray}
\label{E:define-via-PP}
\label{D:Casimir}
\Ec  &\equiv&  \lim_{\eps\to0} \; 
                         \t\half\{\Er(+\eps)+\Er(-\eps)\}
\nonumber\\
     &\equiv& \half\hbar c\mu\cdot \lim_{\eps\to0} \;
                         \t\half\{\z_3(-\t\half+\eps)+\z_3(-\t\half-\eps)\}
\nonumber\\
     &\equiv& \half\hbar c\mu\cdot PP[\z_3(-\t\half+\eps)],
\end{eqnarray}
where the symbol $PP$ stands for taking the principal part. (This
technique yields the ``finite part'' of any meromorphic function that
possesses at worst simple poles.)

The Casimir energy defined in equation (\ref{E:define-via-PP})
depends, in general, on the normalization scale. We keep this scale
dependence to remind us that the that the renormalization programme,
which removes any $\mu$ dependence from the total energy, may
introduce a second finite ambiguity in the Casimir energy. In section
4 we shall study how the Casimir energy varies with this normalization
scale. In section 5 we shall relate the Casimir energy to the one-loop
effective energy, which also depends on the normalization scale. The
difference between the two is finite, $\mu$ independent, and
proportional to the geometric term $C_2$. In particular, the Casimir
and one-loop effective energies agree when $C_2$ vanishes. The total
energy, in the context of bag models, is considered in section 6, and
we shall verify that it is independent of $\mu$.

%\clearpage

%---------------------------------------------------------------
%---------------------------------------------------------------
\section{\bf The role of the normalization scale}
%--------------------------------------------------------------- 
\setcounter{equation}{0}
%---------------------------------------------------------------------

The renormalized Casimir energy defined by equation (\ref{D:Casimir})
generically will depend on the normalization scale $\mu$. This should
not, in fact, be surprising.  As we shall soon see, the Casimir energy
is intimately related to one--loop physics, and the occurrence of
anomalous scale dependence in one--loop field theory calculations is
by now a well understood phenomenon~\cite{Ramond,IZ}.  This anomalous
scaling behaviour manifests itself in two ways: (i) the Casimir energy
may depend on the normalization scale $\mu$; (ii) for conformally
coupled fields, the Casimir energy may fail to scale as the inverse of
the radius of the system. This effect is related to the existence of
the conformal anomaly (trace anomaly). Note however, that the Casimir
energy, in isolation, cannot be measured. What is measurable is the
total energy which includes (renormalized) zero-loop contributions
along with the Casimir energy. If one knew the Lagrangian for the
entire system under study (\eg, see the discussion of bag models later
in this paper) then one would express the total energy in terms of
running coupling constant sand the normalization scale $\mu$. The total
energy is independent of $\mu$.  If the total Lagrangian is unknown,
the Casimir energy still gives the proper geometric dependence for the
oder $\hbar$ part of the total energy. In particular, naive scaling
behaviour of the total energy is violated. The scale $\mu$ should be
interpreted as a scale that summarizes the (unknown) physics
associated with the boundaries, curvature, and masses; it must be
determined experimentally.
 
Consider the effect of a change in the normalization scale
$\mu \to \mu'$. From the definition of the zeta function it is easy to
see that this induces a change $\z_3(s,\mu') = (\mu'/\mu)^{2s} \cdot
\z_3(s,\mu)$, so that $\Er(\eps,\mu') = (\mu'/\mu)^{2\eps} \cdot
\Er(\eps,\mu)$. Now for any analytic function $f(s)$ it is easy to see that
\begin{equation}
PP[f(s)\z(s)] = f(s)\cdot PP[\z(s)] + f'(s)\cdot Res[\z(s)].
\end{equation}
This has the immediate consequence that
\begin{equation}
\Ec(\mu') = \Ec(\mu) -\hbar c \mu \cdot {C_2(\mu) \over (4\pi)^2} \cdot
\ln\left[ {\mu'\over\mu} \right].
\end{equation}
The dependence on the normalization scale is logarithmic, with a
coefficient given by the second Seeley-De Witt coefficient. (The
combination $\mu C_2$ is, despite appearances, independent of the
scale $\mu$.) As is to be expected, this dependence on normalization
scale leads to a breakdown of scale covariance. (It should be noted
that $C_2$ depends on $\int a_2$, and that $a_2$ contains a piece
proportional to the conformal anomaly~\cite{BD}, in fact
$T^\sigma{}_\sigma \propto a_2$, and, for a conformally coupled
theory, $a_2$ {\em is} the conformal anomaly.)
 
Now consider the effect of rescaling the metric and masses: $g_3\to
\kappa^2\cdot g_3$, $m\to \kappa^{-1}\cdot m$. 
This has a simple effect on the eigenvalues of $D_3$, namely:
$\lambda_n\to \kappa^{-2}\cdot\lambda_n$. So for the zeta function
\begin{equation}
\z_3(\kappa^2 g_3; \kappa^{-1} m; s) = \kappa^{2s}\cdot\z_3(g_3;m;s).
\end{equation}
Using the properties of the principal part prescription we find
\begin{equation}
\Ec(\kappa^2\cdot g_3; \kappa^{-1}\cdot m) = {\Ec(g_3;m)\over \kappa}
          -\hbar c\mu \cdot {C_2(g_3;m)\over(4\pi)^2} \cdot
            {\ln \kappa\over\kappa}.
\label{E:scale}
\end{equation}
This is the generalization, allowing for massive particles, of
equation (\ref{E:radius}). It is easy to see that if $\kappa\to\infinity$ then
$\Ec\to 0$, thus the approach to massless particles in Minkowski space
does in fact lead to zero Casimir energy. 
 
To derive equation (\ref{E:radius}) of the introduction, we note that
the radius of the manifold $\kappa^2 \; g_3$ is given by $R(\kappa^2
g_3)=\kappa \; R(g_3)$. Then equation (\ref{E:scale}) may be written
as
\begin{equation}
\Ec(R) = {\hbar c \over R} \cdot
         \{ \eps_0 - \eps_1\cdot\ln(\mu R) \},
\label{E:Radius}
\end{equation}
where
\begin{eqnarray}
\eps_1 &=& {C_2(g_3, \mu=R(g_3){}^{-1})\over (4\pi)^2 \hbar c},
\nonumber\\
\eps_0 &=& \left[{\Ec(g_3,\mu) \cdot R(g_3)\over\hbar c}\right] +
[\eps_1 \; \ln(\mu R(g_3))].
\label{E:eps-0-running}
\end{eqnarray}
Note that $\eps_0$ and $\eps_1$ are independent of the normalization
scale $\mu$. A little thought will show one that $\eps_1$ depends only
on the {\em shape} of the manifold, and are in fact independent of the
radius of the manifold.  The total energy must contain a term with the
same geometric structure as the Casimir energy
\begin{equation}
E_{\mathrm{tot}} = {\hbar c\over R} \; 
\left\{ \eps_0(\mu) - \eps_1 \; \ln(\mu R) \right\} + \dots,
\end{equation}
where now $\eps_0(\mu)$ depends on $\mu$ logarithmically so that
$E_{\mathrm{tot}}$ is independent of the normalization scale. One
might set the scale $\mu$ arbitrarily, and determine the ``running
coupling constant'' $\eps_0$ as a function of $\mu$. In the context of
Casimir energy calculations it is natural to use an alternative
procedure: fix $\eps_0(\mu)$ to have the value determined by equation
(\ref{E:eps-0-running}), and determine $\mu$ experimentally. (This is
completely analogous to the experimental determination of
$\Lambda_{\mathrm{QCD}}$.)

{From} (\ref{E:Radius}) we see that if $C_2(g_3)>0$, then the
Casimir energy has an absolute minimum at $R_{\rm min} = \mu^{-1} \cdot
\exp(1+|\eps_0/ \eps_1|)$, with $E_{min} = -\hbar c|\eps_1|/R_{\rm min}$.
If $C_2(g_3)<0$ then the Casimir energy is unbounded from below,
approaching $E\to-\infinity$ as $R\to0$. (There is now an absolute
maximum at $R_{max}= \mu^{-1} \cdot \exp(1+|\eps_0/ \eps_1|)$ and
$E_{max} = +\hbar c|\eps_1|/R_{\rm max}$.  The {\em sign} of $C_2$ is
thus the determining factor in deciding whether the Casimir effect is
repulsive or attractive for small sizes. If $C_2(g_3)=0$ then an
absolute extremum occurs at $R=\infinity$ and $E=0$.

The appearance of the logarithmic dependence on the radius in
(\ref{E:radius}), (\ref{E:scale}), and (\ref{E:Radius}) is very
striking.  One may quite justifiably ask, would this term not have
been seen in some of the many Casimir energy calculations in the
literature?  The answer is that in very many situations
encountered in the literature $C_2$ vanishes.  Specifically, in flat
3-space, with massless particles, and any collection of infinitely
thin boundaries one has $C_2=0$ (for either Dirichlet or Neumann
boundary conditions).  In particular, considering the case of the
electromagnetic field, any collection of infinitely thin perfect
conductors has $C_2=0$.  To see this, recall $C_2 = A_2 + B_2$.  Now
$A_2 =0$ since we are in flat space.  Further $b_2(y)$ contains
only {\em odd} powers of the second fundamental form.  Infinitely thin
boundaries means that all boundaries consist of two oppositely
oriented faces separated by an infinitesimal distance.  Thus the
second fundamental forms are equal and opposite on the two faces of
each boundary, and consequently the net value of $b_2$ summed over the
two faces of each boundary vanishes.  Thus $B_2=0$, as required.
 
The case of Robin boundary conditions requires extra care. For Robin
boundary conditions ${\curl\phi/\curl\eta(y)} + \psi(y)\phi(y)=0$ on
the boundary. In this case one still has $C_2 = 0$ for thin
boundaries, provided one makes the additional assumption that
$\psi(y+) = -\psi(y-)$. That is, provided $\psi$ is equal and opposite
on the two faces of each thin boundary layer.
 
Some cases where $C_2$ does not vanish have also been discussed in the
literature. These situations have occasioned some rather puzzled
comments which we shall discuss more fully below.
 
%\clearpage
 
%---------------------------------------------------------------
%---------------------------------------------------------------
\section{\bf The one--loop effective action.}
%--------------------------------------------------------------- 
\setcounter{equation}{0}
%---------------------------------------------------------------------

We now consider the relationship between the Casimir energy defined by
(\ref{D:Casimir}) and the one--loop effective energy.  As in the
previous section, we consider an ultrastatic spacetime with $g_4 =
-(dx^0)^2 +g_3$.  To proceed we Wick rotate to imaginary time so that
the Euclidean Laplacian is $D_4 = +\curl_0{}^2 + D_3$.  The heat
kernel then factorizes, $e^{-D_4\mu^{-2}t} = e^{-\curl_0{}^2\mu^{-2}t}
\cdot e^{-D_3\mu^{-2}t}$, so that for the diagonal part of the heat
kernel one has:
\begin{equation}
K_4(x,x,t) = {1\over\sqrt{4\pi \mu^{-2} \; t }} \cdot K_3(x,x,t).
\end{equation}
Now, defining $T=\int dx^0/c = \hbox{``age of the universe''}$, and
applying the Mellin transform (\ref{E:Mellin}) one sees
\begin{equation}
\z_4(s) = {\mu c T \over \sqrt{4\pi}} \cdot
            {\Gamma(s-{1\over2})\over\Gamma(s)}
                \cdot\z_3(s-{\t{1\over2}}).
\end{equation}
Using $E_{\rm eff}\cdot T = S_{\rm eff} = +\half\ln\det D =
-\half\z_4'(0)$, and the known analyticity properties of the zeta function
yields:
\begin{equation}
E_{\rm eff} = \e + \t\half\hbar c\mu \cdot[\psi(1)-\psi(-\t\half)]
                    \cdot{C_2\over(4\pi)^2}.
\end{equation}
Where $\psi(s) = d\ln\Gamma(s)/ds$ is the digamma function.  The
effective energy and Casimir energy differ, but the difference
reflects the inherent renormalization-scheme ambiguity introduced in
the Casimir energy by removing the pole in equation
(\ref{E:regulated}). The unambiguous parts of the effective and
Casimir energies agree, illustrating a remarkably close connection
between zero-point energies and one-loop quantum effects.  Note that
when $C_2=0$, so that the zeta-function regulated Casimir energy is
unambiguous and finite, $E_{\mathrm{eff}} = \Ec$.

There are several variations on the concept of ``vacuum energy'' in
common circulation. One of these is the vacuum--expectation--value of
the integral of the $00$ component of stress energy: $E_{\rm Vacuum} =
\int <0|T_{00}|0>$. This version of the vacuum energy is, in
general, not equal to either one of $\Ec$ or $\Eeff$. However, if one
were to switch off all interactions, so that $T_{00} \to T_{00}^{\rm
Free}$, then an argument, (Presented, \eg, in the review
article~\cite{Review:Casimir}), shows that under rather general
conditions $\Ec = \int <0|T_{00}^{\rm Free}|0>$.  Yet another version
of vacuum energy is obtained by considering the full effective action
in place of the one--loop effective action and its corresponding
effective energy $E_{\eff}^\infinity = \Gamma_{\eff} / T $. Again this
effective energy is quite distinct from the other versions of the
vacuum energy discussed above.  These at least four subtly different
versions of the vacuum energy has unfortunate consequences insofar as
many papers in the literature do not take the appropriate care to make
these distinctions.

%\clearpage
 
%---------------------------------------------------------------
%---------------------------------------------------------------
\section{\bf Comparison with standard results.}
%--------------------------------------------------------------- 
\setcounter{equation}{0}
%---------------------------------------------------------------------

In this section we shall make connections between our formalism and
some of the explicit calculations already available in the
literature. While agreeing with many of those calculations, we report
some subtle differences when considering solid conductors and closely
related aspect of bag models.
 
%---------------------------------
\subsection{Parallel Plates:}
%---------------------------------

Consider a massless scalar field satisfying Dirichlet boundary
conditions confined between two parallel plates of surface area $S$
held a distance $L$ apart. The three dimensional heat kernel is easily
seen to be $K_3(x,x,t) = K_1(x,x,t) / (4\pi\mu^{-2}t)$, which upon
integration over the volume between the plates yields
\begin{equation}
 K_3(t) = {\mu^2 S \over 4\pi t} \cdot K_1(t).
\end{equation}
But $K_1(t)$ is explicitly known in terms of the eigenvalues of the
reduced one dimensional problem $\lambda_n = n^2/L^2$. Evaluation of
the three-dimensional zeta function proceeds in a straightforward
manner
\begin{eqnarray}
\z_3(s)  
&=& 
{\mu^2 S\over\Gamma(s)} 
\int_0^\infinity dt  \cdot t^{s-1} \cdot {
1\over 4\pi t}\cdot \sum_0^\infinity \exp( -t n^2 /\mu^2 L^2 )      
\nonumber\\
&=&
{\mu^2 S\over4\pi} \cdot (\mu L)^{2s-2} \cdot {1\over s-1} \cdot \z_R(2s-2).        
\end{eqnarray}
Here $\z_R$ is the ordinary Riemann zeta function. In taking the limit
$s\to -\half$ one does {\em not} encounter a pole, so the Casimir energy is
simply
\begin{equation}
\Ec(L,S) = -{{1\over12\pi}}\cdot \half \cdot {\hbar c 2\pi S\over L^3}\cdot
\z_R(-3). 
\end{equation}
It is a standard zeta function result that $z_R(-3) = {1\over120}$,
which finally leads to the well-known standard
result~\cite{Review:Casimir}.  This calculation, though trivial, has
expressed some important ideas. The absence of a pole in the
$s\to-\half$ limit can be traced back to the fact that the plates are
{\em flat}. Because the plates are flat the second fundamental form
vanishes ($\gamma=0$), consequently $b_2=0$, and finally $C_2=0$. This
has the additional interesting effect that the flat--plate Casimir
energy is insensitive to the thickness of the plates.

%------------------------------------------------------- 
\subsection{Cylindrical Shells and Spherical Shells:}
%-------------------------------------------------------

For cylindrical and spherical shells $b_2(outside) =-b_2(inside)$, thus
$C_2(net)=0$, and we may safely use simple dimensional arguments to deduce
\begin{eqnarray}
E_{\mathrm{cylinder}} &\propto& {L\over R^2} , 
\nonumber\\
E_{\mathrm{sphere}}   &\propto& {1\over R}.
\label{E:cylinder-sphere}   
\end{eqnarray}
Note that these dimensional analysis results are merely {\em assumed},
not {\em proved}, in the standard analyses of these
problems~\cite{Boyer,Milton-On-Balls,deRaad-Milton}.  It was by no
means clear, in the days before conformal anomalies became a well
understood part of field theory, that there is anything to prove in
deriving (\ref{E:cylinder-sphere}). Fortunately, the naive result
works for thin shells, but as we shall soon see, leads to confusion
when applied to solid conductors. It should be emphasized that the
cancellation of $b_2$ between the inner and outer faces is the
underlying cause of the ``delicate cancellations between internal and
external modes'' noted by many authors~\cite{Review:Casimir}.

%----------------------------------------------------------
\subsection{Solid Cylinders and Solid Spheres:}
%----------------------------------------------------------

For solid conductors the ``delicate cancellations'' alluded to previously
no longer occur. Indeed it is easy to see that
\begin{eqnarray}
C_2(\mu,L,R)_{\mathrm{solid~cylinder}} &\propto& {L\over \mu R^2}
\nonumber\\
C_2(\mu,R)_{\mathrm{solid~sphere}} &\propto& {1\over \mu R} 
\end{eqnarray}
Consequently the Casimir energy possesses a logarithmic dependence on
the radius of these systems. The Casimir energy also depends on the
normalization scale. In regularization schemes such as proper-time
regularization or a mode-sum cut-off the pole associated with $C_2$
manifests itself as an divergent term that depends logarithmically on
the cut--off ~\cite{Kay,Alvarez}.  Such logarithmic divergences have
in fact been encountered in some explicit
calculations~\cite{Milton}. Any term of the form $\ln(R\Lambda)$ may
be re--cast as $\ln(R\mu)+\ln(\Lambda/\mu)$; the $\ln(\lambda/\mu)$
may then be absorbed into a renormalization of some appropriate piece
of the energy, but a term of form $\ln(R\mu)$ always remains in the
renormalized energy (with the $\mu$ dependence compensated by some
other term).

%---------------------------------------------
\subsection{Membranes:}
%---------------------------------------------

We now turn to a very different physical system, that of a membrane.
Membrane theory, as a generalization of string theory, has enjoyed
some recent popularity~\cite{Bars,Kikkawa,Fujikawa}.  Consider a
physical field that is constrained to propagate on the surface of a
closed static membrane. As far as the Casimir effect is concerned,
this is equivalent to considering a 2+1 dimensional spacetime. The
analysis of this paper continue to hold, with the sole exception that
the pole of the zeta function at $s=-\half$ is now proportional to
$C_{3\over2}$. Since $a_{3\over2}$ is automatically zero, this means
that a closed (\ie, boundary-less) membrane automatically has
$C_2=0$. Consequently, zeta-function calculations of the Casimir
effect on any closed membrane are always {\em guaranteed} to not
encounter a pole. This explains the otherwise quite miraculous
cancellation of poles encountered in explicit computations performed
by Sawhill~\cite{Sawhill}.  Open membranes, on the other hand, may
possess poles in the zeta function as $s\to-\half$. The residues of
such poles are, however, tightly constrained.
 
These above comments are also relevant to other physical systems:
consider any field theory that gives rise to domain walls. It is very
easy in such theories to arrange for massless particles to become
trapped on the domain wall. This suggests the interesting possibility
that for suitable choices of parameters and particle content, one may
use the Casimir energy to stabilize small spherical domain walls
against collapse.  Preliminary calculations seem encouraging.
 
At a more general level, the comments of this section imply that the
behaviour of the Casimir effect depends crucially on whether the total
number of spacetime dimensions is even or odd. This will be discussed
more fully when we make some comments on Kaluza--Klein models.

%-------------------------------------------------- 
\subsection{Bag Models:}
%--------------------------------------------------

Another physical situation where the Casimir effect has been of great
importance is in the bag models of QCD~\cite{Bender-Hays,Milton,Baacke}.
As a first approximation, the idea is to treat quarks and gluons as
massless particles confined to the interior of some (3+1)-dimensional
bounded region of spacetime called the bag.  The free quark-gluon
Lagrangian is then augmented by a ``bag Lagrangian'' responsible for
confining the quarks and gluons.
 
The points we wish to make are twofold. First, generically $C_2\neq0$
for these bag models (barring fortuitous cancellations between the
effects of quark and gluon boundary conditions). In cut-off
regularizations of the mode sum this would correspond to the
appearance of a logarithmic divergence, as has indeed been reported by
Milton~\cite{Milton}. In our zeta-function approach the Casimir energy
of the bag includes a $\ln(\mu R)/R$ term. Since we are working with a
model that is supposed to be an approximation to QCD, and since we
have argued that the Casimir energy is related to one--loop effects,
it is natural for the bag models to expect $\mu$ to be related to
$\Lambda_{\rm QCD}$ ($\hbar c\mu \approx \Lambda_{\rm QCD}$).
 
The second point we wish to make concerns the (renormalized) bag
energy.  The {\em total} bag energy depends on the zero-loop bag
energy, {\em plus} the Casimir energy (\ie, one--loop physics), {\em
plus} higher loop effects (presumably small).  One of the great
virtues of the zeta function approach is that it yields an effective
way of calculating the Casimir energy without requiring a detailed
analysis of the renormalization properties of the bag energy. To
extract the structure of the (renormalizable) Bag Lagrangian the
proper time cutoff is more appropriate. In the proper time formalism
\begin{equation}
\Er(\eps) = {\hbar c\mu\over\sqrt{4\pi}} \; 
\int_\eps^\infty  dt \; t^{-3/2} \; {\tr}' (e^{-t D_3 \mu^{-2}}).
\end{equation}
The resulting divergences in the Casimir energy are described by
\begin{equation}
\Er(\eps) \sim 
{C_0\over\eps^2} + {C_\half\over\eps^{3/2}} + {C_1\over\eps} + 
{C_{3/2}\over\eps^{1/2}}+  C_2 \ln\eps + \hbox{finite pieces}.
\end{equation}
Thus the requirement of
renormalizability of the energy implies that the zero-loop bag
energy contains (at a minimum) the following terms
\begin{equation}
E_0= \int_\Omega \sum_0^2  g_n  \; a_n +
     \int_{\curl\Omega} \sum_0^2 h_n \; b_n.
\end{equation}
In flat spacetime this simplifies considerably
\begin{equation}
E_0 = p\cdot V + \sigma\cdot S + 
\int_\bd \left( h_1\; b_1 + h_{3/2}\; b_{3/2} + h_2 b_2 \right).
\end{equation}
Here $p$ is the bag pressure, $\sigma$ is its surface tension, the
parameters $h_1$, $h_{3/2}$ and $h_2$ do not appear to have standard
names.

If we approximate the bag as spherical, we can easily extract the
dependence of these terms on bag radius
\begin{eqnarray}
h_1 \int b_1 &=& F R,
\\
h_{3/2} \int b_{3/2} &=& k,
\\
h_2 \int b_2 &=& h/R.
\end{eqnarray}
Which allows us to write the zero-loop renormalized bag energy as
\begin{equation}
E_0 = p\cdot V + \sigma\cdot S + F R + k + h/R
\end{equation}
It is to be emphasized that these parameters are to be determined by
experiment; they cannot be calculated within the confines of the bag
model.  In principle they would be calculable from the full theory of
QCD.  Adding the one-loop effects (Casimir energy) and defining
$Z=h+\eps_0$ finally yields
\begin{equation}
E_{\mathrm{bag}} = p\cdot V + \sigma\cdot S + F R + k + Z/R - \eps_1 \ln(\mu R)/R.
\end{equation}
The only one of these parameters that is calculable using Casimir
energy techniques is $\eps_1$. In particular, the parameter $Z$ is not
calculable, but rather is to be experimentally determined. The terms
involving $p$ and $\sigma$ are standard. The term involving $F$ has
previously been discussed in the work of Milton~\cite{Milton}. The
offset term $k$ has (to the best of our knowledge) not previously been
discussed. We note in passing that the offset piece $k$ contains a
purely topological piece proportional to the Euler characteristic of
the bag.

%\clearpage
 
%---------------------------------------------------------------
%---------------------------------------------------------------
\section{\bf Applications to Kaluza--Klein theories.}
%---------------------------------------------------------------
\setcounter{equation}{0}
%---------------------------------------------------------------------

In this section we seek to extract some information concerning the
one-loop contributions to the effective four-dimensional cosmological
and Newton constants within the framework of Kaluza--Klein theory.
Calculations along these lines have been carried out, for some
specific simple choices of the internal geometry, in
references~\cite{Applequist-Chodos,Chodos-Meyers,Gilbert-McClain,Gilbert-McClain-Rubin}.
We shall proceed with a bare minimum of assumptions.  Consider a $4+d$
dimensional universe with $d$ compactified dimensions, $\M_{4+d} =
\M_4 \otimes \Omega$. Assume the theory to possess multidimensional cosmological
($\Lambda$) and Newton ($G$) constants.  That is
\begin{equation}
S_{4+d} = \Lambda\cdot\int\sqrt{g_{4+d}}\, d^{4+d}x
+ G^{-1}\cdot\int R_{4+d}\, \sqrt{g_{4+d}}\, d^{4+d}x + \cdots
\end{equation}
Using the product decomposition of spacetime one infers $R_{4+d} = R_4 +
R_d$, so that for the {\em tree--level} four dimensional effective
Cosmological and Newton constants one deduces:
\begin{eqnarray}
\Lambda_{\rm eff} &=& \Lambda\cdot {\mathrm{vol}}(\Omega) + G^{-1}\cdot\int_\Omega
\sqrt{g_d}\, R_d, 
\nonumber\\
G_{\rm eff}^{-1} &=& G^{-1}\cdot {\mathrm{vol}}(\Omega). 
\end{eqnarray}
 
To evaluate the one--loop contributions to $\Lambda_{\rm eff}$ and
$G_{\rm eff}$ one uses the product decomposition of spacetime to deduce a
product decomposition for the diagonal part of the heat kernel
\begin{equation}
K(t) = K_4(t) \cdot K_d(t).
\end{equation}
The asymptotic expansion of the four-dimensional heat kernel may now be
used to obtain an expansion for the zeta function
\begin{equation}
\z_{4+d}(s) = \Sum_0^\infinity {C_n(g_4)\over (4\pi)^2} \cdot
                          {\Gamma(s-2+n)\over\Gamma(s)} \cdot
                           \z_d(s-2+n).
\label{E:reasonable}
\end{equation}
This expansion is a formal one in the ``size'' of the compactified
dimensions. To justify the above expansion consider a ``long
wavelength'' approximation implemented by rescaling the external
dimensions: $g_{4+d,\kappa} = g_{4,\kappa} \oplus g_d = (\kappa^2 g_4)
\oplus g_d$. 
In this situation the heat kernel enjoys the property that
$K_{4+d,\kappa}(t) = K_{4,\kappa}(t) \cdot K_d(t) = K_4(\kappa^{-2} t)
\cdot K_d(t)$. 
Thus the limit $\kappa\to\infty$ allows one to employ the asymptotic
expansion of the heat kernel to obtain an asymptotic expansion for the
multi-dimensional zeta function
\begin{equation}
\z_{4+d,\kappa}(s) = 
\sum_0^N {C_n(g_4)\over(4\pi)^2} \; \kappa^{4-2n} \; 
{\Gamma(s-2+n)\over\Gamma(s)} \;
\zeta_d(s-2+n) + \order(\kappa^{4-2n}).
\end{equation}
By abuse of notation we have rewritten this asymptotic expansion as
the physically more reasonable (\ref{E:reasonable}). Now, recall that
 $C_0=\mu^4\int\sqrt{g_4}\, d^4x$ and $C_1 = k \cdot \int R_4 \,
\sqrt{g_4}\,d^4x$, ($k$ is a constant depending on the
statistics and spins of the elementary particles present in the
theory). This may be used to extract the one-loop corrections to
$\Lambda_\eff$ and $G_\eff$
\begin{eqnarray}
\Lambda_{\eff} &=& 
\Lambda\cdot {\mathrm{vol}}(\Omega) + 
G^{-1}\cdot\int_\Omega\sqrt{g} R_d -
{\mu^4\over2(4\pi)^2} \left\{ \t\half\z'_d(-2) - {\t{3\over4}}\z_d(-2) \right\}. 
\nonumber\\
G_{\eff}^{-1} &=& 
G^{-1}\cdot {\mathrm{vol}}(\Omega)  -
k \cdot {\mu^2\over2(4\pi)^2} \left\{ \z'_d(-1)  - \z_d(-1) \right\}.
\label{E:KK}
\end{eqnarray}
Observe that the zeta functions appearing in the above are guaranteed
to be analytic at all non-positive integers, so that these expressions
are finite as they stand. Further, the value of the zeta function at
non-positive integers is (in principle) known; for example $\z_d(-2) =
2 C_{2+(d/2)} / (4\pi)^{d/2}$, and $\z_d(-1) = - C_{1+(d/2)} /
(4\pi)^{d/2}$.

Without evaluating equation (\ref{E:KK}) in full detail, we may
profitably inquire as to the dependence of $\Lambda_\eff$ and $G_\eff$
on the ``radius'' of the internal dimensions. The major point to be
made is that the case of an odd number of internal dimensions behaves
in a qualitatively different manner form an even number of internal
dimensions. Introducing appropriate constants permits us to write
\begin{eqnarray}
\Lambda_\eff &=& 
a r^d + b r^{d-2} + \left\{ \eps_0 - \eps_1 \; \ln(\mu r) \right\} r^{-4},
\nonumber\\
G_\eff^{-1} &=&
a' r^d  + \left\{ \eps_0' - \eps_1' \; \ln(\mu r) \right\} r^{-2}.
\end{eqnarray}
The dimensionless constants $\eps_1$ and $\eps_1'$ are proportional to
$\z_d(-2)$ and $\z_d(-1)$ respectively. In any odd number of
dimensions (provided the internal manifold has no boundary) these are
guaranteed to vanish. Thus in an odd number of dimensions,
$\Lambda_\eff$ and $G_\eff$ have a simple power-law dependence on the
radius of the compact dimensions. This breaks down however, for any
even number of dimensions where one observes the appearance of
logarithmic dependences on the radius. We expect these logarithms to
have significant effects, but shall postpone further comments to
another paper.

%--------------------------------------------------------------
%\clearpage 
%---------------------------------------------------------------
\section{\bf Conclusion.}
%--------------------------------------------------------------- 
\setcounter{equation}{0}
%---------------------------------------------------------------------

The Casimir energy is a very useful concept, it may be viewed as the
``zero point energy'' of the vacuum, and, from a slightly different
viewpoint, is also intimately related to one--loop physics in the form
of the one--loop Effective energy. In this paper we have exhibited a
unified framework that allows us to regularize and renormalize the
zero point mode sum in a way that is extremely general. Our definition
yields a well behaved finite quantity in many interesting physical
situations: \eg\ in the presence of a background gravitational field,
with massive or massless particles, and in the presence or absence of
boundaries of the space--time manifold. It is hoped that with this
framework in place, it will be possible to perform extensive explicit
calculations.

%---------------------------------------------------------------
\section*{\bf Note added in proof}
%--------------------------------------------------------------- 

After submittal of this paper we were made aware of additional work by
the Manchester
group~\cite{Dowker-Kennedy,Dowker-Banach,Dowker-CQG}. For additional
work on the relevance of the Casimir effect to the stability of
Kaluza--Klein models see
references~\cite{Ordonez-Rubin,Lutken-Ordonez,Birmingham1,Birmingham2}.
In addition we wish to thank Emil Mottola for useful discussions.

%--------------------------------------------------------------
\clearpage
%---------------------------------------------------------------
%---------------------------------------------------------------
\section*{Appendix A}
%---------------------------------------------------------------
\renewcommand{\theequation}{A.\arabic{equation}}
%---------------------------------------------------------------
\setcounter{equation}{0}
%---------------------------------------------------------------------
\centerline{\bf The Seeley--de Witt coefficients.}
%--------------------------------------------------------------- 

The Seeley--de Witt coefficients $a_n(x)$ are independent of the applied
boundary conditions, but the coefficients do depend on the spin of the
field in question.
\begin{equation}
a_0(x) = 1.
\end{equation}
\begin{equation}
a_1(x) = k \cdot R.
\end{equation}
\begin{equation}
a_2(x) = \A (\Weyl)^2 + \B [ (\Ricci)^2 - {\t{1\over3}}R^2 ]
            + \C \nabla^2 R + \D R^2.
\end{equation}

The boundary coefficients $b_n(y)$ depend on the nature of the boundary
conditions imposed. For Dirichlet or  Neumann boundary conditions
\begin{equation}
b_0(y) = 0.
\end{equation}
\begin{equation}
b_{1/2}(y) = \mp {\sqrt{\pi}\over2}.
\end{equation}
\begin{equation}
b_1(y) = {\t 1\over 3} \tr\gamma.
\end{equation}
\begin{equation}
b_{3/2}(y) = a (\tr\gamma)^2 + b \tr(\gamma^2) + c R
\end{equation}
\begin{equation}
b_2(y) = \tilde a (\tr \gamma)^3 + \tilde b (\tr \gamma^2)(\tr\gamma)
       + \tilde c (\tr\gamma^3) + \tilde d (\tr\gamma) R
       + \tilde e \gamma_{ij} R^{ij} +\tilde f \nabla^2(\tr\gamma).
\end{equation}
Where $\gamma$ is the second fundamental form of $\bd$, the boundary
of $\Omega$. The curvatures appearing in $b_n$ are intrinsic
curvatures computed from the induced metric on the boundary. If one
adopts Robin boundary conditions ${\curl\phi\over\curl\eta} +
\psi(y)\phi(y) =0$, then additional terms appear in $b_n$ for $n \geq
1$. Since $\psi$ has the same dimensions as $\gamma$, these extra
terms are of the type exhibited above with $\gamma\mapsto\psi$.
 
%--------------------------------------------------------------- 
%\clearpage 
%---------------------------------------------------------------
\section*{Appendix B}
%---------------------------------------------------------------
\renewcommand{\theequation}{B.\arabic{equation}}
%---------------------------------------------------------------
\setcounter{equation}{0}
%---------------------------------------------------------------------
\centerline{\bf Gamma Function Identities.}
%--------------------------------------------------------------- 
 
We collect some useful Gamma Function identities, see for
instance~\cite{Abramowitz-Stegun}. Take $ n \in \{0,1,2,\cdots\}$:
\begin{equation}
Res[\Gamma(-n+\eps)] = {(-)^n\over n!}.
\end{equation}
\begin{equation}
PP[\Gamma(-n+\eps)] = (-)^n \cdot {\psi(n+1)\over\Gamma(n+1)}
                    = \psi(n+1) \cdot Res[\Gamma(-n+\eps)].
\end{equation}
\begin{equation}
\Gamma(\t\half) = \sqrt{\pi}
\end{equation}
\begin{equation}
\Gamma(-\t\half) = -\sqrt{4\pi}.
\end{equation}
\begin{equation}
\psi(1)= -\gamma.
\end{equation}
\begin{equation}
\psi(n) = -\gamma + \sum_{k=1}^{n-1} {1\over k}.
\end{equation}
\begin{equation}
\psi(\t\half) = -\gamma - 2\ln2.
\end{equation}
\begin{equation}
\psi(\t\half\pm n) = -\gamma - 2\ln2 +2\sum_{k=1}^n {1\over(2k-1)}.
\end{equation}
\begin{equation}
\psi(-\t\half) = -\gamma -2\ln2 +2.
\end{equation}

%--------------------------------------------------------------- 
\clearpage
%---------------------------------------------------------------
%---------------------------------------------------------------
%---------------------------------------------------------------
%---------------------------------------------------------------

%---------------------------------------------------------------
\end{document}